\documentclass[a4paper,11pt]{article}
\pdfoutput=1 

\usepackage{jinstpub} 

\usepackage{lineno}
\usepackage{siunitx}
\newcommand{\pII}{Phase 2 }

\title{\boldmath Current status of the CMS Inner Tracker upgrade for HL-LHC}

\author[a,b]{L. Damenti}

\affiliation[a]{Università degli Studi di Firenze}
\affiliation[b]{INFN Sezione di Firenze}

\emailAdd{lorenzo.damenti@cern.ch}

\abstract{During the High Luminosity programme of the LHC collider (called HL-LHC), planned to start in 2030, the instantaneous luminosity will be increased from \num{\sim 2e34}~\si{cm^{-2}s^{-1}}
to an unprecedented figure of about \num{\sim 7.5e34}~\si{cm^{-2}s^{-1}}. This will allow the Compact Muon Solenoid (CMS) to collect up to \num{\sim 4000}~\si{fb^{-1}} of integrated luminosity over a decade.

In order to cope with the much higher pp-collisions rate, CMS will undergo an extensive improvement known as \pII upgrade: in particular, the silicon tracker system will be entirely replaced to comply with the extremely challenging experimental conditions.

This contribution will review the main upgrades of the CMS Inner Tracker and will present the most relevant design and technological choices. Moreover, the ongoing validation of prototypes and the preparation for the large-scale production will be discussed.}


\collaboration[c]{for the Tracker Group of the CMS Collaboration}

\proceeding{20$^{\text{th}}$ Trento Workshop on Advanced Silicon Radiation Detectors\\
  Feb 4 –6, 2025\\
  FBK, Trento}

\begin{document}
\maketitle
\flushbottom

\section{The High Luminosity LHC program}
\label{sec:intro}
\begin{figure}[h!]
    \centering
    \includegraphics[scale=.35]{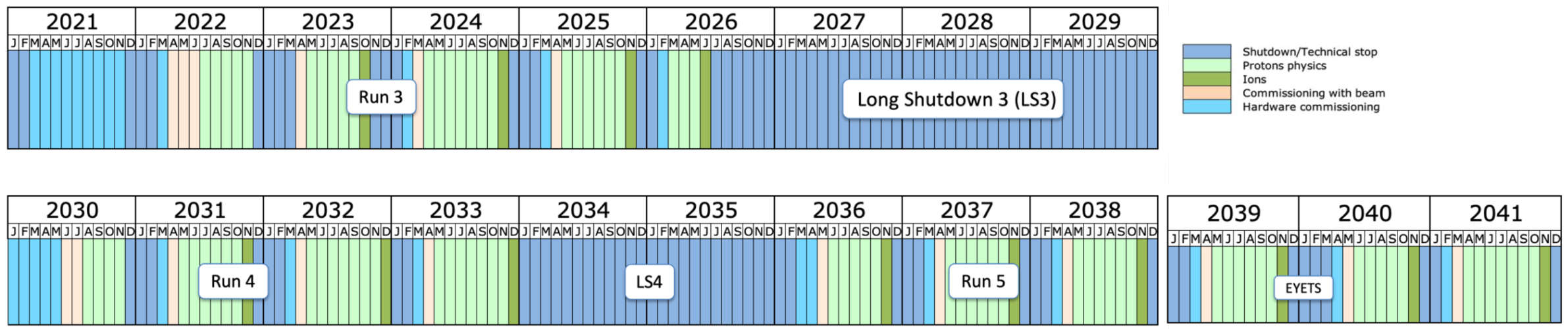}
    \caption{Longer term LHC schedule}
    \label{schedule}
\end{figure}

The High Luminosity program of the LHC, known as HL-LHC, is expected to start in 2030, right after Long Shutdown 3 (see Fig.~\ref{schedule}). In this new era of the Large Hadron Collider, the instantaneous luminosity will be drastically increased.
In the ultimate scenario, it will reach a value of \num{7.5e34}~\si{cm^{-2}s^{-1}}, with an average pileup of 200 interactions per bunch crossing (about three times higher than the average value recorded in 2024). The CMS physics program aims to collect \SI{4000}{fb^{-1}} within 10 years of operation, which is about 10 times more than the current LHC design.

However, experimental operations at high luminosity are extremely challenging. The high pileup, for example, will make trigger and event reconstruction much more difficult. Moreover, radiation damage will be severe, with doses up to \num{\sim 12}~\si{MGy} and fluences of about \num{\sim 2.3e16}~\si{\frac{n_{eq}}{cm^{2}}} in the detector regions closest to the beamline. 

\section{The CMS \pII Tracker}
In order to cope with the demanding new conditions imposed by the HL-LHC program, the Compact Muon Solenoid (CMS)~\cite{a}, one of the four main LHC experiments, will undergo a major upgrade, known as "\pII upgrade".

\begin{figure}[h!]
    \centering
        \includegraphics[scale=0.4]{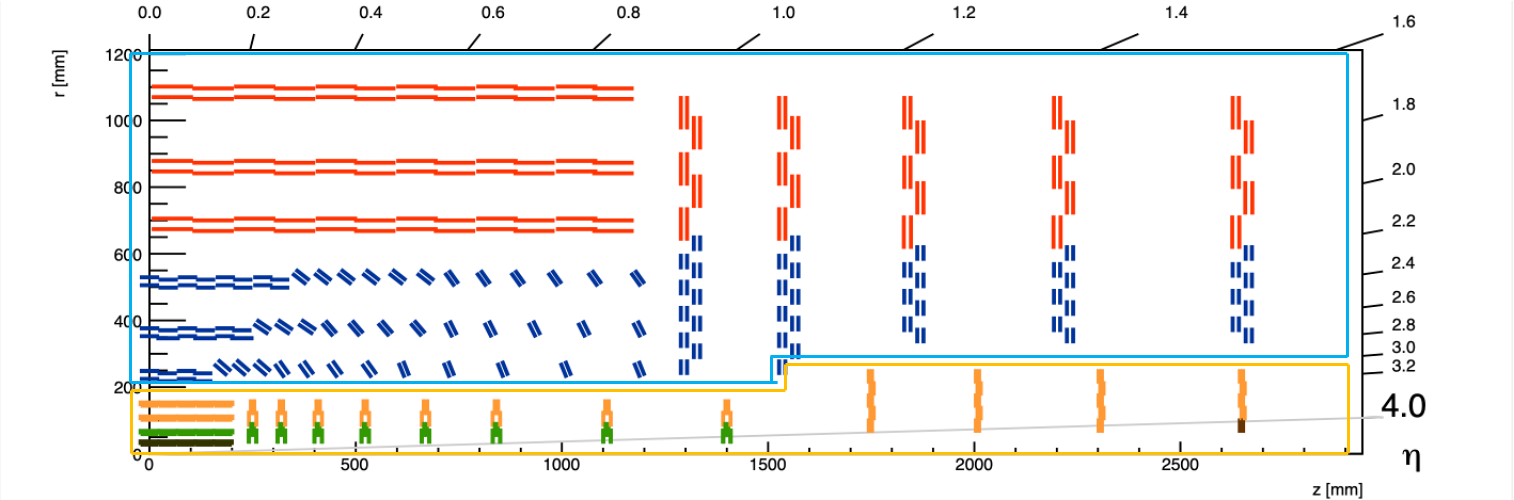}
        \caption{Layout of the CMS \pII tracker. The OT and IT are highlighted, respectively, in blue and yellow. The red layers consist of 2S modules, while the blue layers are instrumented with PS modules. Orange and green layers feature planar sensors, whereas 3D sensor technology will be used in the black layer.}
        \label{trk_layout}
\end{figure}

In particular, the CMS tracker, i.e., the charged particle detector located at the heart of CMS, will be entirely replaced. The \pII tracker~\cite{b} will consist of two parts: the Outer Tracker (OT) and the Inner Tracker (IT). The former will use both strip sensors and macropixels, assembled into two types of modules: 2S (with microstrips) and PS (with macropixels and microstrips), while the latter will feature \num{\sim 2e9} pixels with dimensions of $25 \times 100\ \si{\micro\meter\squared}$. Moreover, the IT will incorporate both 3D and planar sensor technologies. The overall layout of the CMS \pII tracker is shown in Fig.~\ref{trk_layout}.

The CMS \pII tracker features an increased granularity, ensuring that the occupancy remains at the same level as the current tracker (below $0.1\%$) and an extended acceptance up to $|\eta|=4$. 

The following sections will provide a detailed description of the IT system.

\section{The Inner Tracker}
\subsection{Layout}
\begin{figure}
    \centering
    \includegraphics[scale=.6]{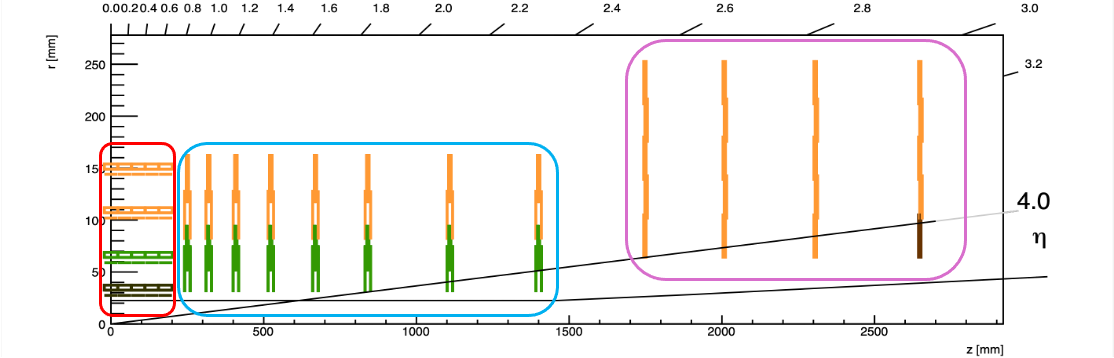}
    \caption{Layout of the CMS \pII Inner Tracker. Different subsystems are highlighted using rectangles of different colours: TBPX in red, TFPX in light blue, and TEPX in magenta. The orange and green layers/disks are instrumented with quad and double planar sensors, respectively, while 3D double modules will be mounted on the black layer.}
    \label{IT_layout}
\end{figure}

As reported in Fig.~\ref{IT_layout}, the CMS \pII Inner Tracker will be formed by three subsystems: Tracker Barrel Pixel (TBPX), Tracker Forward Pixel (TFPX) and Tracker Endcap Pixel (TEPX) and it will be instrumented with more than 3800 hybrid modules. In particular, the IT will feature three different module types: double 3D module (with two 3D sensors and two readout chips), double planar module (with one planar sensor and two readout chips) and quad planar module (with two planar sensors and four readout chips). In general, 3D sensors have a depletion voltage which is 4 times smaller with respect to planar ones. This translates into lower power consumption, which leads to a higher margin with respect to thermal runaway\cite{c}. However, this technology will be deployed only in the first layer of the TBPX, as its production processes are more complex compared to those of planar sensors, resulting in higher costs and lower production yields.

\subsection{Mechanics}
The mechanics of the Inner Tracker upgrade have no direct mechanical connection to the beampipe. This design facilitates IT extraction, making both maintenance during technical stops and potential replacement of the innermost TBPX layer and TFPX disk more straightforward.

A key feature of the IT mechanics is its overall lightweight structure, achieved through the use of carbon foam and carbon fiber. Additionally, the implementation of an evaporative $\mathrm{CO_2}$ cooling system further reduces weight, thanks to efficiency of the approach that translates onto the small diameter of cooling pipes (<~\num{2}~\si{mm}).

\subsection{Powering and Read-Out scheme}
\begin{figure}[!h]
    \centering
    \includegraphics[scale=0.47]{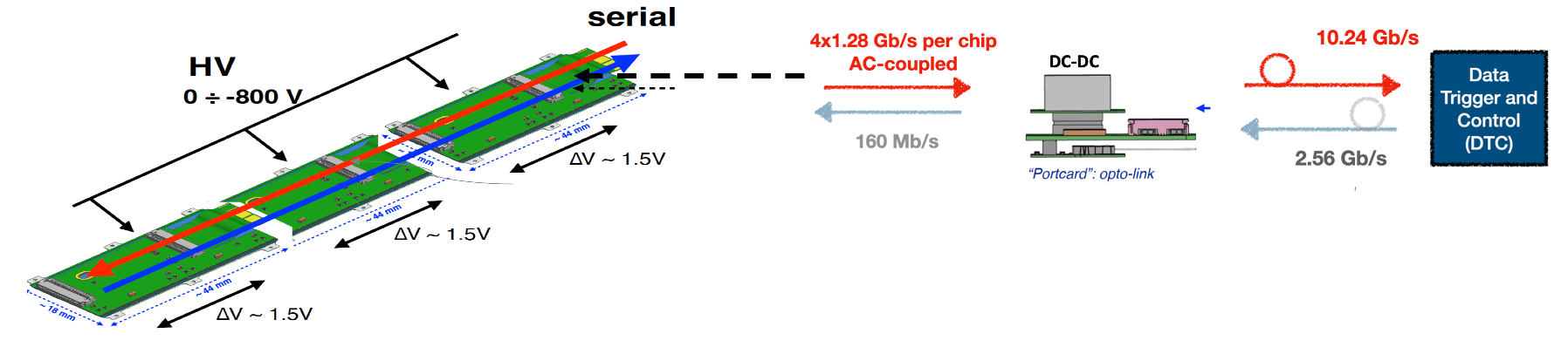}
    \caption{Left: An example of a three-module serial power chain. Right: A schematic representation of the readout system.}
    \label{Pow_Rout}
\end{figure}

The key innovation of the Inner Tracker lies in its module powering scheme. The modules will be arranged into 576 serial power chains, as serial powering is the only viable solution for supplying a 50 kW detector while maintaining a low material budget. The material budget, which represents the amount of material a particle encounters as it traverses the detector, is a critical factor in minimizing multiple scattering and preserving tracking precision. Figure~\ref{Pow_Rout} (left) illustrates a schematic of a serial power chain consisting of three modules. Within each module, the chips are powered in parallel, and this same parallel scheme is used to distribute high voltage to each module.
Regarding data readout, e-links will be used to transmit data from the module to a Low Power GigaBit Transceiver (lpGBT) at an uplink speed of \num{1.28}~\si{Gbps} and a downlink speed of \num{160}~\si{Mbps}. Up to three lpGBTs will be integrated into a custom opto board known as portcard. Optical fibers will ensure the connection between each portcard and the Data, Trigger, and Control (DTC) boards. A schematic of the full readout chain is shown in Fig.~\ref{Pow_Rout} (right).

\subsection{Inner Tracker Modules}
\begin{figure}[!h]
    \centering
    \includegraphics[scale=.4]{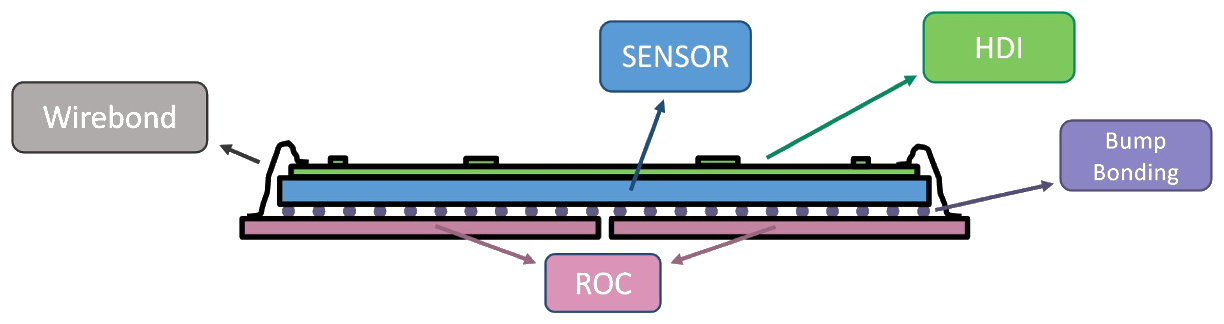}
    \caption{Sketch of the general structure of an IT module. The HDI is glued on top of the flip-chip, i.e. the assembly of the ROC and sensor, which is made using the bump-bonding technique. aluminium wires, knonw as wirebonds, connect the HDI to the ROC. The module is protected against sparks using parylene coating.}
    \label{mod_sketch}
\end{figure}

The general structure of a typical Inner Tracker module is sketched in Fig.~\ref{mod_sketch}.
The electrical connection between each pixel of the sensor and the elementary cell of the readout chip (ROC) is achieved using a tin-silver alloy (SnAg) bump. This hybridization technique, known as bump-bonding, is carried out by various vendors, such as IZM, Advafab, and Micross. The High-Density Interconnector (HDI), that hosts ancillary passive electronics and ensures power distribution and data readout, is glued on top of the ROC-sensor assembly, also referred to as flip-chip. The connection between the HDI and the ROC is established via aluminium wires known as wirebonds. Additionally, all IT modules are coated with parylene, which provides protection against high-voltage discharges (sparks).

\subsubsection{CMS Read-out Chip}
The Read-out Chip for the CMS \pII upgrade of the Inner Tracker, known as CROC (CMS Read-Out Chip), is the CMS version of a ROC developed through a joint effort between CMS and ATLAS within the RD53 collaboration~\cite{d}. In particular, the CROC is a \si{65}{nm} CMOS Application-Specific Integrated Circuit (ASIC) designed to withstand radiation doses of up to \SI{10}{ MGy} and featuring a linear front-end with an adjustable threshold in the 1000 electrons range. These characteristics are crucial for operation in a hostile radiation environment. On the other hand, its fine granularity ($432 \times 336$ $50 \times \SI{50}{\mu m^2}$ cells over a surface of $2.2 \times 1.9\ \si{cm^2}$) is key for physics performance.

\subsubsection{Sensors}
\begin{figure}
    \centering
    \includegraphics[scale=0.6]{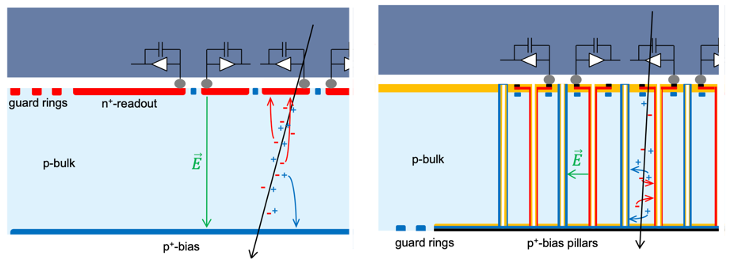}
    \caption{Sketch illustrating the main differences between planar (left) and 3D (right) technologies.}
    \label{planar_VS_3D}
\end{figure}

The Inner Tracker modules will be equipped with two different silicon sensor technologies: planar and 3D sensors. The former feature thin superficial electrodes implanted on the surface of the substrate, while the latter have cylindrical electrodes that penetrate deeply into the substrate (see Fig.~\ref{planar_VS_3D}). The main properties of the two different technologies are summarized in Table~\ref{tab}. 

\begin{table}[ht]
    \centering
    \resizebox{\textwidth}{!}{  
    \begin{tabular}{|l|l|}
        \hline
        3D sensors & Planar Sensors \\
        \hline
        \SI{150}{\mu m} active thickness + \SI{100}{\mu m} support & \SI{150}{\mu m} active thickness and no support \\
        No elongated pixels, \SI{175}{\mu m} periphery & Elongated pixels in the interchip region, \SI{450}{\mu m} periphery \\
        temporary metal for wafer-level IV qualification & no bias dot (IV only from guard ring or test structures) \\
        $V_{\mathrm{dep}}<10V$, $V_{\mathrm{bd}}>V_{\mathrm{dep}}+35V$ (fresh) $V_{\mathrm{bd}}>200V$ (irrad) & $V_{\mathrm{bd}}>350V$ (fresh) $V_{\mathrm{bd}}>600V$ (irrad) \\
        hit efficiency > 96\% (normal incidence) &
        hit efficiency > 99\%\\
        \hline
    \end{tabular}
    }   
    \caption{Table summarizing the main differences between the 3D and planar sensors.}
    \label{tab}
\end{table}

A planar sensor consists of two pixel matrices each matching the readout chip layout. To eliminate the inefficiency gap between the matrices, elongated pixels are introduced at the periphery. In contrast, a 3D sensor features a single pixel matrix with the same number of pixels but without elongated ones.
This distinction influences the design of the double module assembly: a double planar module consists of two ROCs and one planar sensor, whereas a double 3D module is composed of two ROCs and two sensors. The inefficiency region between 3D matrices does not affect detector performance, as it is covered by modules from the subsequent layer

It is also worth mentioning that the hit efficiency of 3D sensors is 96\% with normal incidence, due to an intrinsic inefficiency of the technology: if a particle crosses a column of the sensor, it cannot be detected. However, this inefficiency does not affect the detector performance, as particles typically cross the tracker with an angle.

Throughout the years, many test beam campaigns have been carried out to validate the performance of both 3D and planar sensors after irradiation. In particular, two main parameters have been defined for these studies: efficiency ($\epsilon_{\mathrm{hit}}$) and acceptance ($\alpha$), which are defined as follows:  $\epsilon_{\mathrm{hit}} = \frac{N_{\mathrm{tracks}}^{\mathrm{observed}}}{N_{\mathrm{tracks}}^{\mathrm{total}}}$,  $\alpha = 1 -  \frac{N_{\mathrm{masked}}^{\mathrm{pixels}}}{N_{\mathrm{total}}^{\mathrm{pixels}}}$. 
In the definition of acceptance, $N_{\mathrm{masked}}^{\mathrm{pixels}}$ represents the number of pixels in the sensor matrix that have been masked because they are either stuck or noisy. A pixel is considered stuck if its occupancy is below a given threshold when a high charge (i.e., $10000$ electrons) is injected into the front-end preamplifier by means of the charge injection circuitry that the ROC is equipped with. On the other hand, if a pixel has an occupancy higher than a given threshold without any charge injection, it is labeled as noisy. Typically, the occupancy threshold is the same for both definitions and is set to $10^{-6}$. The DAQ software, called Phase2 Acquisition and Control Framework (Ph2ACF)~\cite{e}, has different procedures for performing test injections into the ROC front-end electronics, thereby determining the total number of stuck and noisy pixels.  

\begin{figure}
    \centering
    \includegraphics[scale=0.5]{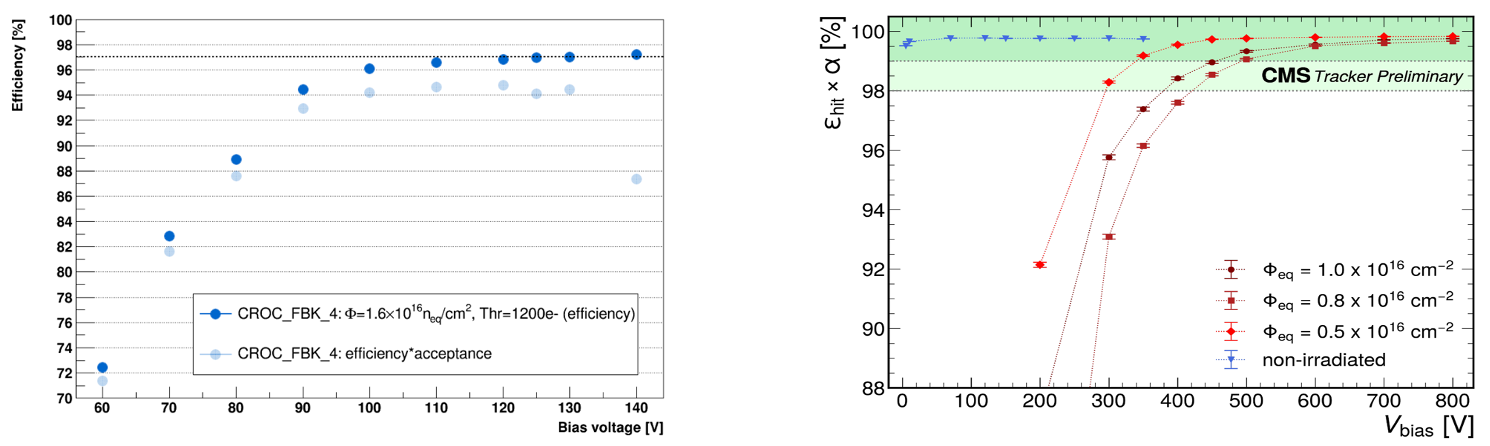}
    \caption{Left: Results for an irradiated 3D sensor with a fluence of \num{1.6e16}~\si{\frac{n_{eq}}{cm^{2}}}. With a depletion voltage of $110V$, the efficiency exceeds 96\%, meeting the requirements. Right: Results for an irradiated planar sensor with a non-uniform fluence ranging from \num{0.5e16}~\si{\frac{n_{eq}}{cm^{2}}} to \num{1e16}~\si{\frac{n_{eq}}{cm^{2}}}. With a depletion voltage of $600V$, the efficiency exceeds 99\%, meeting the requirements.}
    \label{TB_res}
\end{figure}

Figure \ref{TB_res} shows some results obtained during a test beam campaign with irradiated sensors~\cite{f}~\cite{g}. In particular, for the 3D sensors, the fluence was approximately \num{1.6e16}~\si{\frac{n_{eq}}{cm^{2}}}, while for the planar sensors, it was not uniform, ranging from \num{0.5e16}~\si{\frac{n_{eq}}{cm^{2}}} to \num{1e16}~\si{\frac{n_{eq}}{cm^{2}}}. For both sensor types, the results met the requirements, with an efficiency above 96\% for the 3D sensor ($V_{\mathrm{depletion}}=110V$) and above 99\% for the planar sensor ($V_{\mathrm{depletion}}=600V$).  

\subsubsection{Module types, production and testing}
\begin{figure}[!h]
    \centering
    \includegraphics[scale=0.5]{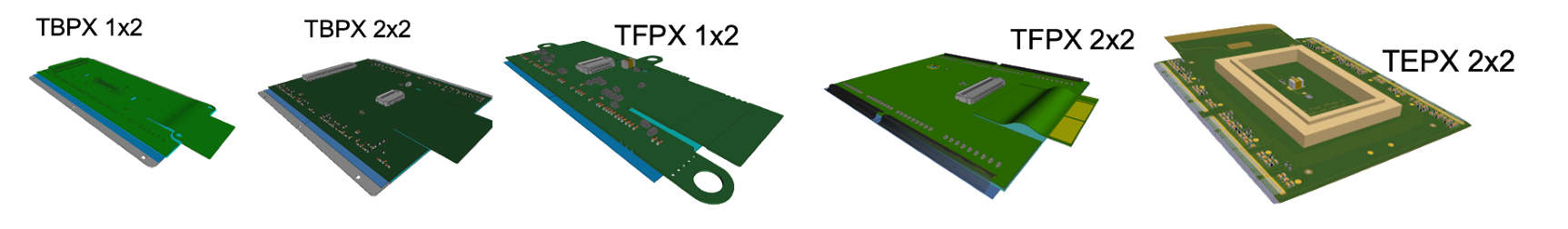}
    \caption{Different module types considering the IT subsystems.}
    \label{ModTypes}
\end{figure}

Considering the different subsections of the IT and the various sensor technologies, modules can be classified into three distinct groups:
\begin{itemize}
    \item \textbf{Barrel Modules:} Planar double (two ROCs and one sensor), Planar Quad (four ROCs and two sensors) and 3D double (two ROCs and two sensors)
    \item \textbf{Forward Modules:} Planar Double, Planar Quad
    \item \textbf{Endcap Modules:} Planar Quad
\end{itemize}
It is worth noting that barrel modules include a cooling plate as a fixation interface to the mechanical structure. Different module variants are shown in Fig.~\ref{ModTypes}.

\begin{figure}[!h]
    \centering
    \includegraphics[scale=0.6]{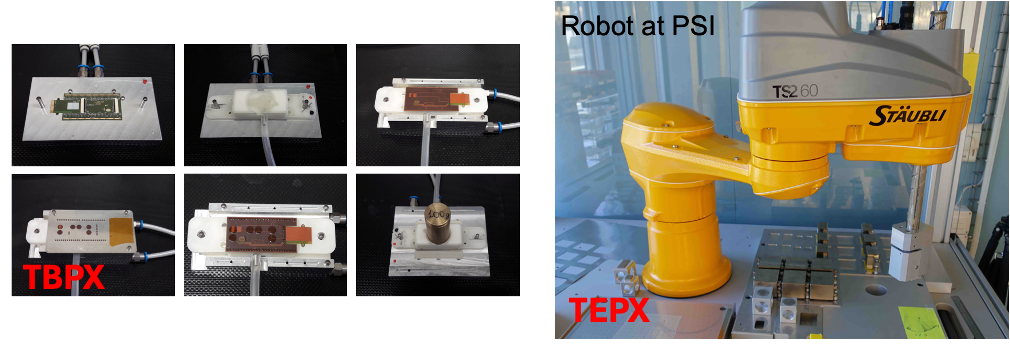}
    \caption{Different tools used for module assembly. Manual jigs for TBPX (left) and robotica arm for TEPX (right).}
    \label{tools}
\end{figure}

The different types of modules are assembled in various centers using different tools. For example, the TBPX assembly centers perform manual assembly using jigs, while the TEPX and TFPX module assembly centers employ a fully automated robotic process. Figure~\ref{tools} shows TBPX jigs (left) and a TEPX robotic arm (right).

After module assembly, several tests must be performed to assess its quality, including: visual inspection, I-V curve measurement, module tuning and performance evaluation, bump-bonding test, and thermal cycling.

\begin{figure}
    \centering
    \includegraphics[scale=0.65]{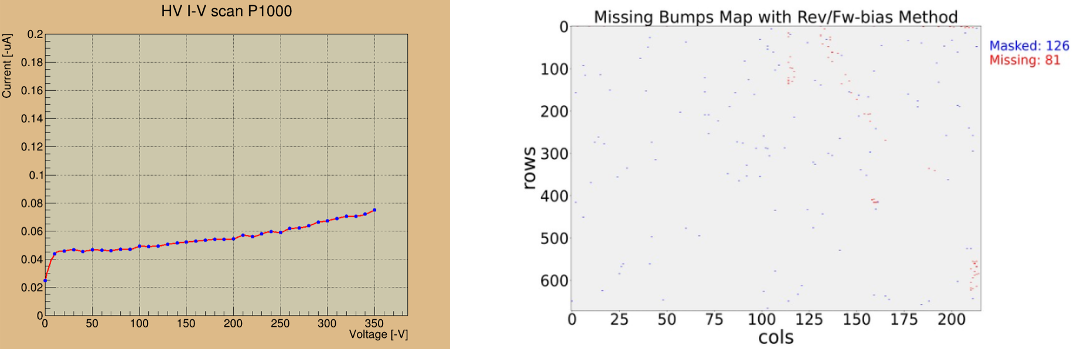}
    \caption{Results for I-V Curve test (left) and Bumb-bonding test (right). Since the number of missing bumbs is below the threshold of 600, this can be considered as a good chip.}
    \label{tests}
\end{figure}

The I-V curve test is essential for characterizing the flip-chip. Specifically, it is used to measure the current drawn by the sensor and to determine when it reaches the breakdown voltage. An example of an I-V curve is reported in Fig.~\ref{tests} (left)~\cite{h}.

Module tuning refers to the process of adjusting the threshold of each chip to approximately $1000$ electrons. This is carried out using dedicated calibrations in the Ph2ACF. The DAQ software also allows for evaluating the module's noise performance. Typically, if a flip-chip has a number of noisy pixels that is smaller than 1\% of the total number of pixels (i.e. smaller than $\sim 1400$), it is considered as a good chip.

Another crucial test after module assembly is the bump-bonding test. As the name suggests, this test is intended to verify the quality of the bump bonds in terms of electrical connection on each chip. Generally, if the number of missing or disconnected bumps is below 600, the chip passes this test. An example of a result from a bump-bonding test is shown in Fig.~\ref{tests} (right)~\cite{i}.

All of the tests described above are performed in a coldbox that can accommodate up to eight modules.

\subsection{Current status}
Most of the components for the CMS \pII upgrade of the Inner Tracker are either in production or in the pre-production phase. Regarding module assembly, the final module production is set to begin soon. In particular, TBPX and TFPX assembly centers have just received the final versions of both the CROC and the HDI, allowing them to start assembling the first pre-production modules. This step is crucial for validating the assembly strategy with the final components. Regarding TEPX, assembly centers have already built four pre-production modules.

On the testing side, all procedures are well defined, but full automation is still in progress to ensure a high-throughput production process.

Finally, module integration is scheduled to begin in October 2025 for TFPX, November 2025 for TEPX, and January 2026 for TBPX. 

\section{Summary}
The High-Luminosity LHC program will define a completely new experimental environment, requiring the full replacement of the CMS Tracker to cope with the resulting harsh conditions. 

The \pII Inner Tracker (IT) is subdivided into distinct regions (TBPX, TFPX, and TEPX) and features a lightweight mechanical structure along with a serial powering scheme. To meet performance and radiation requirements, the IT will incorporate two types of sensor technologies: 3D sensors and planar sensors. The data readout will be managed by the CROC chip, a 65 nm CMOS ASIC specifically developed for this upgrade. Module assembly will take place across multiple specialized centers, and strict quality control procedures are in place to ensure consistency and reliability. 

Currently, most of the components for the Inner Tracker are either in production or in the pre-production phase. Final module production is now starting. All testing procedures have been clearly defined, and full automation for high-throughput production is ongoing. The integration of modules is scheduled to begin on October 25 for TFPX, November 25 for TEPX, and January 26 for TBPX.


\end{document}